# Redesigning Fronthaul for Next-Generation Networks: Beyond Baseband Samples and Point-to-Point Links

Jingchu Liu, Shugong Xu, Sheng Zhou, and Zhisheng Niu

# Abstract

The fronthaul (FH) is an indispensable enabler for 5G networks. However, the classical fronthauling method demands for large bandwidth, low latency, and tight synchronization from the transport network, and only allows for point-to-point logical topology. This greatly limits the usage of FH in many 5G scenarios. In this paper, we introduce a new perspective to understand and design FH for next-generation wireless access. We allow the renovated FH to transport information other than time-domain I/Q samples and to support logical topologies beyond point-to-point links. In this way, different function splitting schemes could be incorporated into the radio access network to satisfy the bandwidth and latency requirements of ultra-dense networks, control/data (C/D) decoupling architectures, and delay-sensitive communications. At the same time, massive cooperation and device-centric networking could be effectively enabled with point-to-multi-point FH transportation. We analyze three unique design requirements for the renovated FH, including the ability to handle various payload traffic, support different logical topology, and provide differentiated latency guarantee. Following this analysis, we propose a reference architecture for designing the renovated FH. The required functionalities are categorized into four logical layers and realized using novel technologies such as decoupled synchronization layer, packet switching, and session-based control. We also discuss some important future research issues.

# 1 Introduction

In recent years, cellular networks have witnessed a tremendous surge in data traffic, which is largely driven by the wide-spread adoption of smart devices such as smart phones and tablets [1]. While this trend will most likely continue in the foreseeable future, new challenges are also emerging with the proliferation of machine-type communications and real-time cloud services. In response to these predicted challenges, next-generation networks are envisioned to provide 1000x capacity, 100x data-rate, and 1ms latency compared with 4G LTE systems [2]. The enabling technologies for such supreme performance are expected to include massive MIMO, ultra-dense network (UDN), as well as high-frequency spectrum.

Fronthaul (FH) is an important enabler for the deployment of these technologies in 5G

networks. The term FH[1] has its root in the distributed base station (BS) architecture, in which the processing functions of a BS are split into two entities: the Remote Radio Unit (RRU) which takes charge of radio processing and digital-analogue conversion near the antennas, and the Baseband Unit (BBU) which handles digital baseband processing at another location. The classical form of FH refers to the point-to-point (P2P) link that transports time-domain complex baseband radio (a.k.a I/Q) samples between the corresponding RRU and BBU. Recently, FH also finds usage in the novel Centralized Radio Access Network (C-RAN) architecture [3]. In C-RAN, time-domain I/Q samples are aggregated from scattered antenna sites to a central office for uplink (UL) processing or sent out in the opposite direction after downlink (DL) processing.

The dominant physical transmission technology for classical FH is digital radio-over-fiber (D-RoF). Although there are also several competing technologies such as analog radio-over-fiber, the longer transportation range in C-RAN gives D-RoF great advantages due to its low signal deterioration. Various specifications have been formed to support the inter-operability between FH products from different manufacturers. For example, the Common Public Radio Interface (CPRI) specification [4] covers the layer 1 and layer 2 of FH. Its scope includes the physical topology, line data rates, framing format, etc.

Although classical FH has been widely adopted in distributed BS and C-RAN, it will nevertheless face serious challenges in face of 5G networks:

1) **Massive FH bandwidth requirements:** the bandwidth requirements of a classical FH link is proportional to the product of radio bandwidth, number of antennas, and quantization resolution. To put this into perspective, a typical 20MHz 4G LTE eNodeB with 8 antennas requires around 10Gbps FH bandwidth on UL or DL. Since the area density of antennas and the communications bandwidth are both expected to be enormous in 5G networks, the demand for FH bandwidth will become even more significant. Nevertheless, state-of-art compression techniques can only achieve at most 3x compression rate [5]. The most straightforward option for physical transport is dark fiber, in which one fiber core can only carry one FH link. However, this will result in enormous fiber resource consumption, making this scheme realistic only to operators with abundant fiber resources or in scenarios where fiber deployment is cheap. Another option is to multiplex FH links into a single fiber core using wave-division multiplexing (WDM), but WDM modules are still much more expensive than black & white modules.

2) **Stringent latency constraints:** wireless signal processing often have stringent latency constraints. For example, the LTE Hybrid Automatic Repeat reQuest (HARQ) process leaves a latency budget of 3ms for the decoding of each radio sub-frame. Because sub-frames need to be first transported from RRHs to BBUs before being processed, the transportation latency should also be counted into this latency budget. Excluding the portion necessary for signal processing (about 2.5ms), there are only 300 to 500us left for FH transportation, ruling out any switching-based technologies that will incur excessive latency. Moreover, some 5G communication scenarios demand for even stricter latency constraints: for sub-ms wireless access, the latency budget for FH transportation will be so small that any long-range transportation may be prohibited.

3) **Tight synchronization requirements:** synchronization is essential for radio

---

[1] This type of links got "front" in its name since it is closer to the network edge compared with backhaul (BH), which connects different BSs as well as BSs and core-network elements.

communication systems. But many RRHs cannot generate accurate clock by themselves, either because GPS receivers are too expensive to be integrated into RRH or due to satellite signal blockage in indoor environments. For this reason, FH must deliver synchronization information from BBUs to RRHs. In CPRI links, clock information is carried in the waveform (pulse edges) of the transported signal. But the underlying network infrastructure may introduce jitter to the waveform, leaving degraded communication performance. The importance of tight synchronization will be even greater in 5G networks because of the massive cooperation between access nodes. If cooperating access nodes have frequency offset, their transmitted signals will overlap at UE and cannot be separated and individually compensated, causing distorted beam-forming pattern and degraded performance.

In this paper, we propose a renovation for classical FH to address the above challenges. The renovated FH can transport intermediate-processing information beyond time-domain I/Q samples. Also, it supports logical topologies beyond P2P links. The renovated FH can seamlessly incorporate different function splitting schemes and logical topologies, and can enable a number of key 5G concepts. We provide detailed analysis on three unique design requirements for the renovated FH, i.e. 1) handling various payload traffic, 2) supporting flexible logical topology, and 3) providing differentiated latency guarantees. To facilitate efficient realization of the renovated FH network, we introduce a layered reference architecture and discuss how the layers could be realized using technologies such as decoupled synchronization, packet switching, and session-based management. Promising future research issues are also listed.

The rest of the paper is organized as follows. In Section 2, we propose our renovation for classical FH and discuss how the renovated FH can enable some key 5G concepts. In Section 3, we analyze the three fundamental design requirements for the renovated FH. We then propose a layered reference architecture for realizing the renovated FH and introduce the enabling technologies in Section 4. After that, we discuss some important future research issues in Section 5. The article is concluded in Section 6.

# 2 Renovating FH

The classical understanding for FH is a P2P link between a pair of RRH and BBU for transporting time-domain I/Q samples. In this section, we renovate this concept in two ways in order to address the challenges in 5G networks: 1) **FH should transport intermediate-processing information other than time-domain I/Q samples**; 2) **FH shall support not only P2P transportation but also point-to-multi-point networking**. We also illustrate how these new features could enable some promising 5G concepts.

## 2.1 Beyond Time-Domain I/Q Fronthauling

Recent research on BS function splitting revealed that allowing the transportation of intermediate-processing information other than time-domain I/Q samples will help eliminate the bandwidth and latency bottlenecks of classical FH. Function splitting concerns the split of signal processing functions among different entities [6]. The classical RRH-BBU split is an extreme case, with minimal processing at RRHs. In contrast, alternative function splitting schemes place processing functions such as FFT/iFFT, MIMO pre-coding/detection, modulation/demodulation, or even the whole physical layer (PHY) stack at RRHs. A tradeoff between computational and networking resource can be achieved by using different function splitting schemes [7]. The optimal tradeoff point should be decided considering the different resources availability and constraints of specific scenarios. Analysis shows that the FH bandwidth requirements of alternative splitting schemes can be two orders of magnitude lower than that of the classical scheme, and the requirement on latency can also be relaxed [8].

The rationale behind these benefits can be explained as the following: signal processing functions are in a sense appending (extracting) redundant information to (from) the communications signals in order to combat channel deteriorations. For example, modulation maps code-words (6 bits for 64QAM) to complex constellation points (30 bits) so as to combat noises. Placing some of the processing functions at RRHs will reduce the amount of redundant information which needs to be transported, and therefore will reduce the FH bandwidth requirements. As for latency, by moving latency-sensitive processing functions to remote sites, the fronthauling (FHing) latency will be no longer in the total time budget, thus the latency requirements can also be relaxed.

Since different function splitting schemes will result in different FHing payload, FH should be allowed to transport not only time-domain I/Q samples but also other intermediate-processing information. Also, now that processing functions can be flexibly placed, the classical definition for RRH and BBU should also be generalized: RRHs should handle some extra processing functions, and BBUs need to handle less processing functions.

## 2.2 From P2P FHing to FH Networking

Classical FH is in essence a logical P2P link, even though the underlying transport topology may be stars, rings or chains. However, recent developments in cooperative communications put forward the need for point-to-multi-point logical topology. Massive cooperation is a main element in the technology evolution of RAN [9]: the information from (to) multiple RRHs is jointly processed to mitigate interferences or to increase cell throughputs. Typical cooperative processing schemes include joint transmission (JT), joint reception (JR), coordinated scheduling (CS), and coordinated beamforming (CB). As the next-generation networks will become much denser, it is highly possible that cooperative processing become prevalent in 5G networks.

Centralization is a possible way to implement cooperative processing. The processing information from multiple RRHs is aggregated through FH to one BBU as in Figure 1 (a). In this way, the information of cooperating cells can be exchanged inside the BBU. When full centralization is too expensive, neighboring RRHs could form a cooperation cluster and the processing information only needs to be aggregated within the cluster. In such a case, information must also be exchanged between clusters to facilitate communications on cluster edges. One way to exchange information between clusters is to establish FH links from one RRH to multiple BBUs as in Figure 1 (b). For this case, FFT/iFFT may need to be placed at RRHs so as the resource elements used for cooperative communication can be extracted and send to the cooperating cluster(s). Another way is to forward processing information from the BBU of one cluster to that of another cluster as in Figure 1 (c) [2].

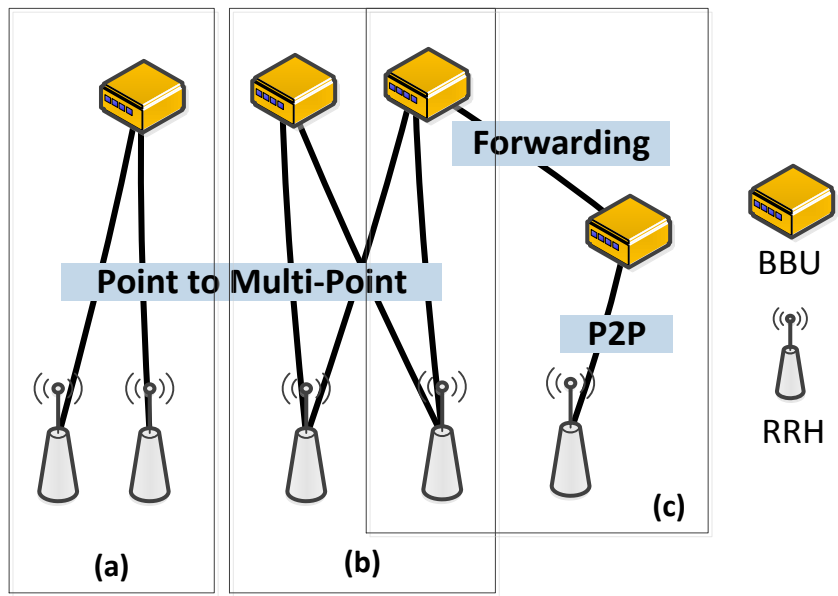

Figure 1 Possible logical topologies for FH networking.

## 2.3 Enabling Key 5G Concepts

### 2.3.1 C/D Decoupling Architectures

The Air-interface C/D decoupling architecture as proposed in hyper-cellular network [10] and phantom cell [11] is a key concept for managing dense small cells in 5G networks. In this architecture, physical control channels are only transmitted by control-plane large cells, giving data-plane small cells chances to hibernate when there is no data transmission. In addition, the large control coverage of control cells also allows for better mobility management.

The renovated FH can support function splitting and facilitate effective deployment of massive small data cells. The enormous FH bandwidth requirements can be reduce by placing pre-processing functions at remote sites or satisfied by utilizing wireless FH link on the last hop; besides, the sparseness and burstiness of small cell traffic will be absorbed by statistical multiplexing, improving the utilization ratio of FH wireline resources. In both ways, the implementation flexibility of UDNs will be greatly improved in areas with limited fiber resource. Also, C/D decoupling inherently implies different FH payloads for control and data cells: control cells generate more control-channel information, while data cells generate more data-channel information. As illustrated in Figure 2, this heterogeneous FHing demands can be satisfied more efficiently (compared with classical FH) by using FH links that are respectively matched to the FH traffic patterns and delay requirements of control and data cells.

---

[2] Forwarding information between BBUs on FH links may look similar to exchanging information between BSs through the X2 interface. However, because the main functionality of FH, i.e. networking intermediate-processing information, is different from that of BH, the realization of their network interface shall be vastly different.

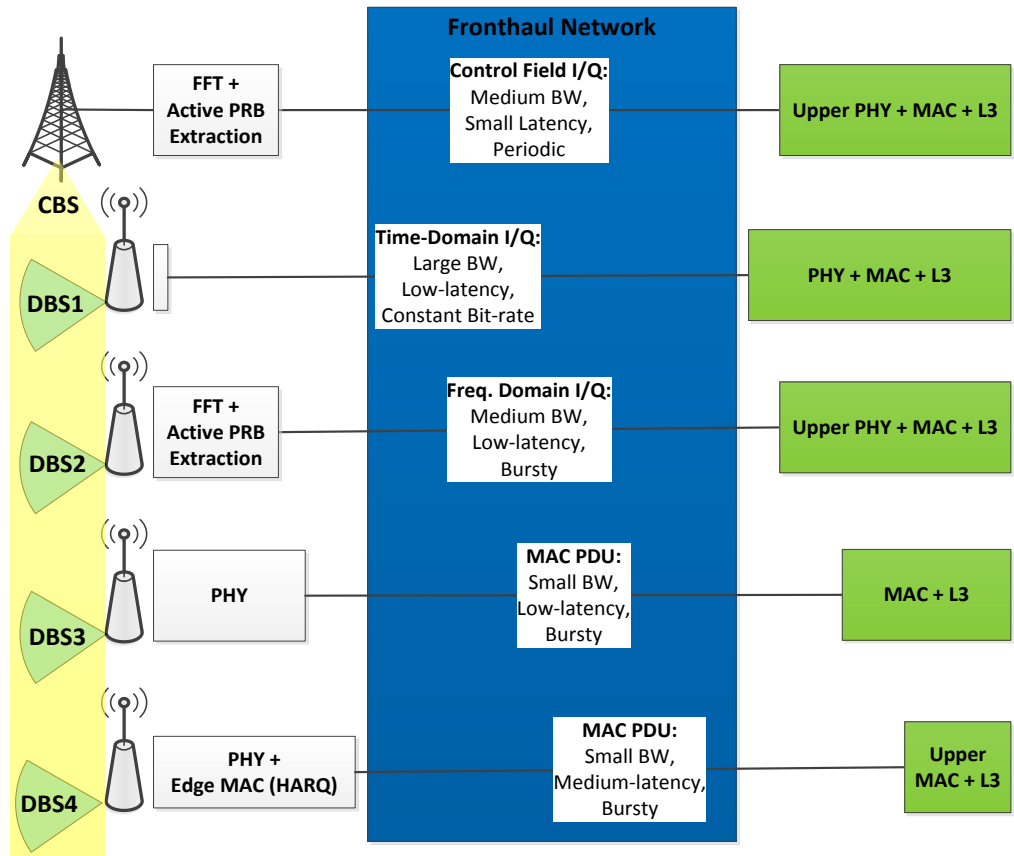

Figure 2 FH networking for C/D decoupled architecture and massive small cells. CBS and DBS can adopt different function splitting schemes (see function names in silver and green boxes) and will result in FH payload with different bandwidth, latency, and traffic pattern. (CBS: control base station, DBS: data base station, PRB: physical resource block, PHY: physical layer, MAC: medium access control layer, L3: layer 3 or network layer, PDU: packet data unit, BW: bandwidth)

## 2.3.2 Device-Centric Communications

Device-centric communications will be one disruptive technology direction for 5G networks [12]. This technology blurs the concept of "cell" and breaks away with the previous cellular communication paradigm, by which devices are connected to only one radio node at any particular point of time. In device-centric communications, devices will simultaneously connect to multiple access nodes. In response, the role of FH should also be altered from "transporting information for cells" to "networking information for devices". As illustrated in Figure 3 (orange), the intermediate-processing information of a user need to be transported simultaneously from multiple RRHs to one BBU for joint processing. Also, different users may send information to different BBUs. Moreover, the set of connected RRHs may change as users move around, thus the set of corresponding FH links need to be adapted accordingly.

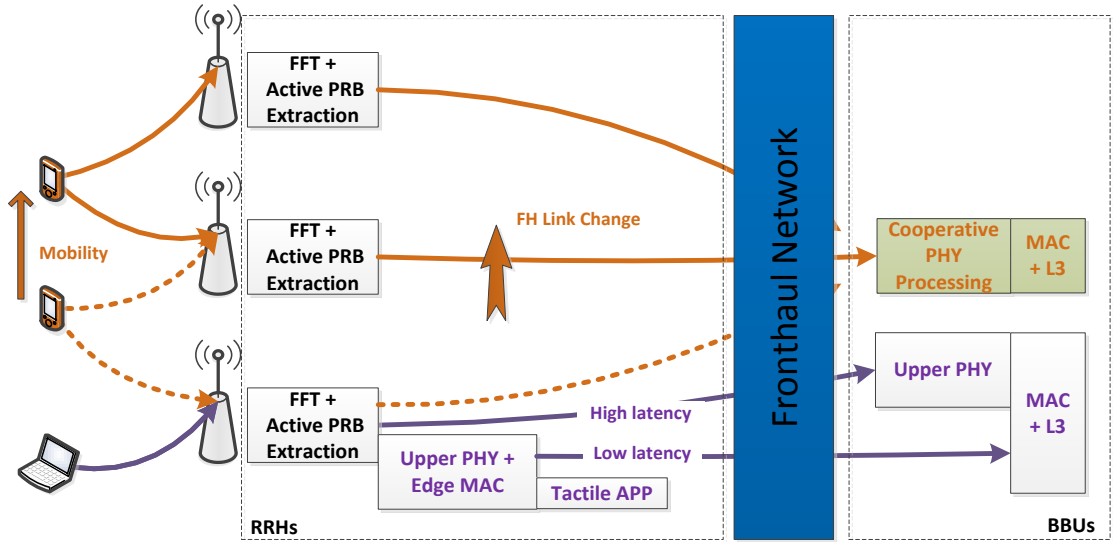

Figure 3 FH networking for device-centric (orange) and latency-sensitive communications (purple) (PRB: physical resource block, PHY: physical layer, MAC: medium access control layer, L3: layer 3 or network layer)

## 2.3.3 Latency-Sensitive Communications

Latency will be one of the key concerns for the applications of 5G networks. Although applications with normal delay requirements, such as common mobile Internet services, will still be needed, there will also be new applications like tactile Internet [13], which demands sharply shorter access latency (about 1ms). The renovated FH can support these diverse latency requirements as in Figure 3 (purple). For latency-sensitive applications, the whole processing stack should be placed at RRHs in order to avoid the FH transportation latency; otherwise, low-latency FH links must be provisioned to ensure the latency constraint is not violated. In contrast, normal-latency applications may utilize FH links with looser latency guarantee. In more common scenarios, the latency requirements from different applications may differ significantly, and the renovated FH will provide a range of latency guarantee options for the resulting FH payloads.

# 3 Design Requirements

In this section, we analyze the three design requirements that are unique to the renovated FH. These requirements include 1) handling various payload traffic, 2) supporting flexible networking topology, and 3) providing differentiated latency guarantee.

## 3.1 Various Payload Traffic

Different function splitting schemes will introduce various bandwidth requirements for FH

network. As can be seen in Figure 4, the baseline classical splitting scheme (time-domain I/Q FHing) consumes the most FH bandwidth. In comparison, the bandwidth requirements of alternative split schemes can be much lower:

1) Low-Pass-Filtered I/Q FHing: Time-domain I/Q samples are low-pass filtered before FH transportation. The blank guard band can be filtered out, roughly halving the FH bandwidth requirements.
2) Resource Element Extraction: All cell processing functions (e.g. FFT/iFFT) are place at RRHs. Only the I/Q samples on active resource elements are extracted for FHing. In such a scheme, the instantaneous FHing rate is proportional to the actual radio resource usage, allowing the FH bandwidth requirements to further fall off in lightly loaded cells.
3) Modulation Bits FHing: MIMO precoding/detection and modulation/demodulation functions are placed at RRH, thus the intermediate-processing information to be transported is actually modulation information bits. Since modulation bits are much more compact representations than I/Q samples, the FH bandwidth requirement is often at least one orders of magnitude lower than baseline.

Note that, these examples just illustrates the basic concept and are far from exhaustive. Other system configuration and function splitting schemes may result in different results.

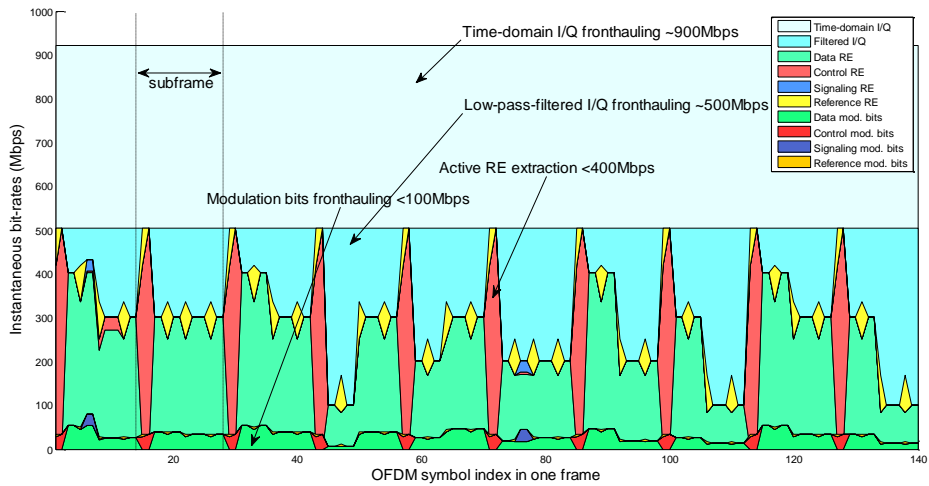

Figure 4 Instantaneous fronthauling bit-rates of different function splitting schemes. Simulation parameters: 20MHz single cell, 10 UE, DL SISO transmission with link adaptation, and sub-frame-level channel and traffic variation.

Aside from bandwidth requirements, the FH traffic patterns of different splitting schemes will also manifest different level of randomness and periodicity (see Figure 4). Randomness is a consequence of cell load variations on one hand: the amount of radio resources under use varies as users come and leave, resulting in time-varying bandwidth requirements. On the other hand, it is also resulted from the time-varying usage of modulation and coding schemes (MCS) in response to channel fluctuations. Meanwhile, periodicity is mainly due to the transportation of periodic control signals like physical downlink control channel (PDCCH), physical random access channel (PRACH), as well as other overhead information.

These diverse bandwidth requirements and traffic patterns will render the traditional constant bit-rate FHing technique, which only offers limited bandwidth and traffic pattern options, inefficient. For this reason, the renovated FH ought to provide sufficient flexibility to efficiently transport payloads with various bandwidth and traffic patterns.

## 3.2  Flexible Networking Topology

The logical topology of FH will be diversified in 5G networks. As mentioned previously, centralized cooperative processing requires FH network to aggregate (distribute) information from (to) multiple RRHs to one BBU or transport information between BBUs. Besides, there may also be the need to transport information among RRHs when cooperative processing functions are placed at RRHs. Moreover, when the function splitting scheme are dynamically changed as in [7], the FH networking topology need to be adapted accordingly.

Aside from the support for different type of logical topologies, an additional design requirement is fine-grained topology management. There are two possible scenarios where the logical topology of FH links needs to be managed on a sub-cell granularity. Firstly, when different cells are cooperating by exchanging information between RRHs, the information produced by control and signaling processing functions may still need to be transported to BBU simultaneously. In this case, the RRH-RRH and RRH-BBU links need to be managed separately. Secondly, the management granularity may need to be down to device-level in device-centric communications: cooperative processing information for cell-edge users can be transported to the BBU, while the processing functions of other non-cooperative users can be executed locally at RRHs.

The renovated FH should have the ability to provide a much diverse collection of logical topologies, and provide fined-grained routing granularity for its payloads. Also, since the routing strategies will be tightly coupled with the selection of function splitting schemes, FH should provide the interface to enable joint design of function splitting and FH transportation.

## 3.3  Differentiated Latency Guarantee

Differentiated latency guarantee is needed for both the evolution of current networks and the development of 5G networks. In current networks, different function splitting schemes may result in different latency requirements. Take LTE for example, if all processing functions that are below HARQ are conducted at RRHs, the latency requirement for FH can be relaxed to 10ms level (due to PRACH latency constraint) [14]. For 5G networks, the demand for differentiated latency guarantee mainly comes from new applications. At one extreme, sub-ms wireless access is needed by future real-time applications such as tactile Internet. In such as case, either the transport latency on FH should be minimized, e.g. to below 100us, or the whole processing stack should be place at RRH to avoid FH transportation. At the other extreme, there are also applications for which latency is not too much a concern. A typical example is delay-tolerant machine-type communications. In such a case, the latency requirements on FH can be correspondingly relaxed. The relaxed latency requirements makes payload scheduling and switching possible. Also, transport network with inherently large delay can be used in scenarios where no low-latency infrastructure is available. In between of these two extreme cases, applications with intermediate latency requirements may also exist. FH should have the ability to provision FH links with differentiated latency guarantee for different applications.

# 4  Realization Aspects

In this section, we discuss the realization aspects of the renovated FH. To facilitate discussion, we first propose a reference architecture for the renovated FH. As illustrated in Figure 5, the physical layout of the proposed architecture is constructed of RRHs, BBUs, and FH switches. They are inter-connected using physical links and form certain network topology, such as rings or chains.

The functionality of these physical network elements can be categorized into four logical layers. The foundation is the synchronization layer for distributing timing and clock references; in the middle are the payload layer for payload forwarding and the control layer which oversees payload forwarding and timing distribution; on the top sits the session layer which presents logical FH links for applications. Next, we discuss some enabling technologies for these layers.

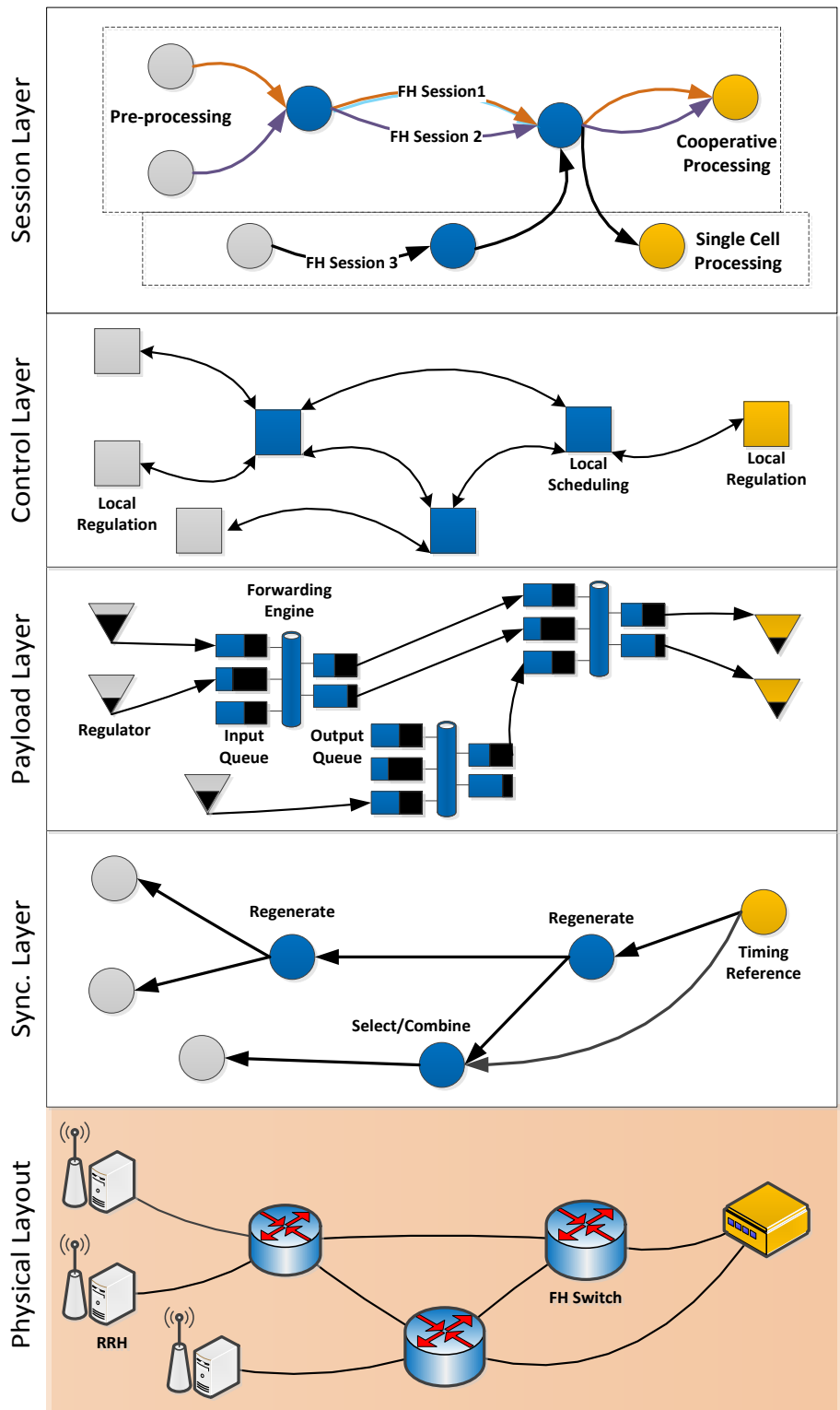

Figure 5 Reference architecture

## 4.1 Decoupled Synchronization

As mentioned in Section 1, synchronization is essential for radio communication systems. In Classical FH, synchronization signals are transported on FH links together with I/Q payloads. This coupled mode works well on P2P links, but will create many ambiguities as the network topology become more complex. To address this difficulty, the synchronization functionality in the proposed architecture is decoupled from payload transportation: the logical topology for timing and clock distribution is decided separately with payload transportation. Specifically, 1) RRHs are slave nodes and synchronized to the signals from upstream FH nodes. 2) BBUs receive synchronization signals either from other nodes or from external timing sources. 3) FH switches handle the selection, combination, and regeneration of timing and clock signals. External timing sources can also be attached to FH switches and serve as inputs. These processes are handled by dedicated circuitry like phase-locked loop (PLL) and specially-designed message-based timing algorithms. Separate synchronization layers like this can also be found in other designs such as Synchronous Ethernet and IEEE 1588 [15].

## 4.2 Packet Switching

FH payload is transported in the FH network in form of packets. As illustrated in the payload layer of Figure 5, the data streams generated at RHHs or BBUs are first traffic-shaped by the regulator through buffering and framing, forming the transportation payloads. Each frame is then attached with a packet header which indicates its FH link number, link counter number, etc. The resulting packet is then handed to lower-layer protocols for physical transportation. On the receiver side of the physical link, the FH switch first buffers the packet in the input buffer and extract its header information. The local controller then processes the header and decides the scheduling and forwarding policy for this packet. The packet is then moved to the output buffers along with the payloads from other FH links, and transported to the next-hop FH switch. Once the packet has arrived at the destination end equipment, it is again buffered at the regulator, striped of its header, and recovered into payload data stream for further signal processing.

Packet-switched transportation has several advantages. Firstly, variable length packet can effectively handle the various payload traffic resulting from different function splitting schemes. Secondly, packetized payloads can be separately scheduled to increase the bandwidth utilization of physical link under diverse payload bandwidth and traffic patterns, and to provide differentiated latency guarantee. Thirdly, packets can be individually forwarded, forming the desired logical topology. Lastly, the FH payload can be dynamically switched to redundant physical paths under link or node failure, increasing the overall resilience of the network. Note packet-based transportation of time-domain I/Q samples is being investigated by IEEE 1904 Work Group under the name of radio-(CPRI-) over-Ethernet. In contrast, our proposal also supports the transportation of information beyond I/Q samples and logical topologies other than P2P links.

## 4.3 Session-based Control

The capacity of the FH network is presented in the form of FH sessions. A FH session stands for a stream of FH payloads between a pair of end equipment and having certain bandwidth and latency guarantee. The payloads in the same session share the same physical path in the network. Each session is set up and tore down separately in the network by the control layer. The control layer calculates the best transportation path and configures the local control entities along the path.

This session-based control is realized using virtual-circuit switching, which establishes virtual connections in a packet-switched network. Connect-oriented switching has a number of advantages over its connectionless counterparts, such as better bandwidth and latency guarantee, lower switching overhead, and higher switching speed. This kind of technology has been widely used in major protocols like Multiprotocol Label Switching (MPLS). These features are all favorable for FH transportation, since FH payloads may have different bandwidth and latency requirements, but also demand high switching capacity and low switching delay.

# 5 Future Research Issues

In this section, we discuss some important future research issues for the renovated FH.

## 5.1 Overhead Analysis and Reduction

The renovated FH network provides great flexibility over the payload traffic, topology, and latency guarantee. However, along with the great flexibility also comes increased overhead. Firstly, the packet headers will thus reduce the overall bandwidth efficiency. Secondly, header processing and payload scheduling will introduce additional transportation latency and require more computational resources. Thirdly, fine-grained session management will bring more negotiation and signaling overhead.

These overhead should be thoroughly quantified and analyzed in the future in order to identify fundamental design tradeoffs, which can be used for overhead reduction. For example, shortening the header may reduce the amount of control information that can be carried in each packet and hurt flexibility, but it will increase bandwidth efficiency and speed up the control processing pipeline. Likewise, restrictions on available payload length options may reduce the bandwidth efficiency, but will provide better latency guarantee. Also, managing FH sessions in a group will reduce flexibility but will reduce the overall signaling overhead at the same time. Efficient implementation will have to search for an optimal tradeoff point.

## 5.2 Software-Defined Networking

Software-defined networking (SDN) is a novel paradigm to enhance the efficiency, flexibility, and management of networks. This concept can also be introduced into the renovated FH. Note in the reference architecture, the control layer does not mandate a distributed implementation. A centralized controller could be introduced into this layer to coordinate payload scheduling and forwarding as well as timing distribution based on global network information. However, in order to introduce SDN into FH networks, the common issues with SDN must be addressed. For example, scalability is a major concern in any architectures with centralized control. The central controller may be overloaded by massive control signaling demands and complex control algorithms, resulting in large latency or packet drop. This issue calls for novel control-plane designs such as hierarchical architecture. Besides, centralized control also facilitates software-defined network operation: applications can access the capabilities of the proposed architecture by calling application interfaces (APIs) on the central controller. But the APIs must be designed according to the special needs of FH networking.

## 5.3 Heterogeneous FH Links

The available link technologies for FH will continue to evolve and blossom in the future. Optical modules will have higher rate and lower price thanks to the development of technologies like silicon photonics. Wireless link technologies such as mmWave and free space optics (FSO) is also likely to become mature enough to transport FH payloads as well as control and synchronization signals. Both types of link technologies have their own pros and cons. Therefore, it is important for the renovated FH to utilized heterogeneous FH links in a synergetic manner. For example, wireless links could provide last-hop FH access, while optical transport can be used to aggregate last-hop access links.

These different network segments may employ different protocols, and thus segment gateways need to be designed to transparently deliver payloads across the FH network. Another issue is the non-uniform link QoS among segments. For example, public networks often have less bandwidth and larger latency than private networks. To provide uniform QoS guarantees along the FH link, performance monitoring, reporting, and negotiation capabilities must be designed at segment gateways, and the QoS differences of different segments should be considered by the FH when scheduling and forwarding payloads. The feasibility and best practice of synergetic operations of heterogeneous FH links require further investigation.

## 5.4 Networking-Processing Co-design

In traditional RAN, the performance of wireless communication is decoupled with computational and wireline resources. Each BS is equipped with enough computational resources to handle signal processing on its own, even in peak hours. And the information exchange demand between different BSs and between BS and the core network is minimized since all signal redundancy is removed through local processing.

However, the renovated FH allows for the co-design of processing and networking. This is because the processing information can now be transported to arbitrary places for processing, allowing for the tradeoff between computational, wireline, and radio resources. Possible tradeoffs include processing consolidation (wireline-, computation+)[3], FH compression and function splitting (computation-, wireline+), centralized cooperative processing (computation-, wireline-, radio+), wireless FH (radio-, computation-, wireline+). Preliminary examples of such trade-offs have emerged in the form of fully centralized processing in C-RAN. However, more alternative forms of tradeoff should be investigated to address the practical challenges of cloud-based radio access, such as high FH bandwidth requirements and energy consumption.

---
[3] Note "+" means the tradeoff is in favor of this kind of resource, while "-" means the tradeoff act against it. If not mentioned, then this kind of resource is not affect by the tradeoff.

# 6 Conclusion

In this paper, we renovate the classical FH to address the challenges in 5G networks. The renovated FH can transport intermediate-processing information beyond time-domain I/Q samples, and allows for logical topologies other than P2P links. In this way, different function splitting schemes and logical topologies could be employed to enable key 5G concepts without violating the bandwidth and latency constraints. With respect to the renovated FH, we highlight the three unique design requirements, i.e. 1) handling various payload traffic, 2) supporting flexible logical topology, and 3) providing differentiated latency guarantees. We also provide a layered reference architecture for realizing the renovated FH and discuss key enabling technologies and important future research issues.

# 7 Acknowledgment

This work is sponsored in part by the National Basic Research Program of China (No. 2012CB316001), and the National Natural Science Foundation of China (NSFC) under grant No. 61201191 and 61401250, the Creative Research Groups of NSFC (No. 61321061), the Sino-Finnish Joint Research Program of NSFC (No. 61461136004), and the Intel Collaborative Research Institute for Mobile Networking and Computing.

# 8 Biographies

JINGCHU LIU [S] (liu-jc12@mails.tsinghua.edu.cn) received his B.S. degree in Electronic Engineering from Tsinghua University, China, in 2012. He is currently a PhD student at the Department of Electronic Engineering, Tsinghua University. His research interests include cloud-based wireless networks, network data analytics, and green wireless communications.

SHUGONG XU [SM] (shugong.xu@intel.com) received his Ph.D. degree from Huazhong University of Science and Technology in 1996. He is currently the Director of Intel Collaborative

Research Institute for Mobile Networking and Computing (ICRI-MNC). Before he took this role, he was Principal Investigator of ICRI-MNC and co-directing the research programs for this new institute after he joined Intel Labs in Sep 2013. Prior to that, he was a research director and principal scientist at the Communication Technologies Laboratory, Huawei Technologies. Among his responsibilities at Huawei, Shugong founded and directed Huawei's green radio research program GREAT. He was also the Chief Scientist and lead for the China National 863 project on End-to-End Energy Efficient Networks. Prior to joining Huawei in 2008, he was with Sharp Laboratories of America as a senior research scientist. Shugong published more than 60 peer-reviewed research papers in top international conferences and journals. One of his most referenced paper has over 1200 Google Scholar citations, in which the findings were among the major triggers for the research and standardization of IEEE 802.11S. Shugong has over 20 US patents granted. Some of these technologies have been adopted in international standards including IEEE 802.11, 3GPP LTE and DLNA. His recent research interests include mobile networking and computing, next generation wireless communication platform, network intelligence and SDN/NFV, etc.

SHENG ZHOU [M] (sheng.zhou@tsinghua.edu.cn) received his B.S. and Ph.D. degrees in Electronic Engineering from Tsinghua University, China, in 2005 and 2011, respectively. He is currently an assistant professor of Electronic Engineering Department, Tsinghua University. From January to June 2010, he was a visiting student at Wireless System Lab, Electrical Engineering Department, Stanford University, CA, USA. From November 2014 to January 2015, he was a visiting researcher in Central Research Lab of Hitachi Ltd., Japan. His research interests include cross-layer design for multiple antenna systems, cooperative transmission in cellular systems, and green wireless communications.

ZHISHENG NIU [F] (niuzhs@tsinghua.edu.cn) graduated from Beijing Jiaotong University, China, in 1985, and got his M.E. and D.E. degrees from Toyohashi University of Technology, Japan, in 1989 and 1992, respectively. During 1992-94, he worked for Fujitsu Laboratories Ltd., Japan, and in 1994 joined with Tsinghua University, Beijing, China, where he is now a professor at the Department of Electronic Engineering and deputy dean of the School of Information Science and Technology. He is also a guest chair professor of Shandong University, China. His major research interests include queueing theory, traffic engineering, mobile Internet, radio resource management of wireless networks, and green communication and networks. He is now a fellow of both IEEE and IEICE, a distinguished lecturer (2012-15) and Chair of Emerging Technology Committee (2014-15) of IEEE Communication Society, and a distinguished lecturer (2014-16) of IEEE Vehicular Technologies Society.